\documentclass[prd,showpacs,amsmath,amssymb,superscriptaddress,floatfix,nofootinbib,10pt]{revtex4}
\usepackage{times}
\usepackage{amssymb,amsbsy,amsmath,amsfonts}
\usepackage{graphicx}
\usepackage{float}
\usepackage{color}
\usepackage{morefloats}
\usepackage{rotating}
\usepackage{srcltx}
\usepackage{slashed}
\usepackage{multirow}
\usepackage{verbatim}
\usepackage{hyperref}
\usepackage{tabularx}
\usepackage{bm}
\allowdisplaybreaks[4]

\usepackage{bbding}
\usepackage{threeparttable}

\newcommand{\PreserveBackslash}[1]{\let\temp=\\#1\let\\=\temp}
\newcolumntype{C}[1]{>{\PreserveBackslash\centering}p{#1}}
\newcolumntype{R}[1]{>{\PreserveBackslash\raggedleft}p{#1}}
\newcolumntype{L}[1]{>{\PreserveBackslash\raggedright}p{#1}}

\begin{document}

\title{Test of the hyperon-nucleon interaction within leading order covariant chiral effective field theory}

\author{Jing Song}
\affiliation{School of Physics, Beihang University, Beijing, 102206, China}

\author{Zhi-Wei Liu}
\affiliation{School of Physics, Beihang University, Beijing, 102206, China}

\author{Kai-Wen Li}
\email[]{kaiwen.li@buaa.edu.cn}
\affiliation{Medical Management Department, CAS Ion Medical Technology Co., Ltd., Beijing 100190, China}
\affiliation{Beijing Advanced Innovation Center for Big Data-Based Precision Medicine, School of Medicine and Engineering, Beihang University, Key Laboratory of Big Data-Based Precision Medicine (Beihang University), Ministry of Industry and Information Technology, Beijing, 100191, China}
\affiliation{School of Physics, Beihang University, Beijing, 102206, China}

\author{Li-Sheng Geng}
\email[]{lisheng.geng@buaa.edu.cn}
\affiliation{School of Physics, Beihang University, Beijing, 102206, China}
\affiliation{Beijing Advanced Innovation Center for Big Data-Based Precision Medicine, School of Medicine and Engineering, Beihang University, Key Laboratory of Big Data-Based Precision Medicine (Beihang University), Ministry of Industry and Information Technology, Beijing, 100191, China}
\affiliation{Beijing Key Laboratory of Advanced Nuclear Materials and Physics, Beihang University, Beijing, 102206, China}
\affiliation{School of Physics and Microelectronics, Zhengzhou University, Zhengzhou, Henan, 450001, China}

\begin{abstract}
Motivated by the recent experimental measurements of differential cross sections of the $\Sigma^{-}p$ elastic scattering in the momentum range of $470$ to $850$ MeV$/c$ by the J-PARC E$40$ experiment, we extend our previous studies of $S=-1$ hyperon-nucleon interactions to relatively higher energies up to $900$ MeV$/c$ for both the coupled-channel  $\Lambda p\rightarrow(\Lambda p, \Sigma^{+}n, \Sigma^{0}p)$, $\Sigma^{-}p\rightarrow(\Lambda n, \Sigma^{0}n, \Sigma^{-}p)$ and single-channel $\Sigma^{+}p\rightarrow\Sigma^{+}p$ reactions. We show that although the leading order covariant chiral effective field theory is only constrained by the low energy data, it can describe the high energy data reasonably well, in particular, the J-PARC E40 differential cross sections.
The predicted cusp structure close to the $\Sigma N$ threshold in the $\Lambda p\to \Lambda p$ reaction agrees with the latest ALICE observation as well as with the results of the  next-to-leading order heavy baryon chiral effective theory. On the other hand, the comparison with the latest CLAS data on the $\Lambda p$ cross sections between 0.9 and 2.0 GeV$/c$ clearly indicates the need of higher order chiral potentials for such high momenta. This is also the case for the latest J-PARC data on the $\Sigma p \rightarrow \Lambda n$ differential cross sections. Nevertheless, even for these cases, the predictions are  in qualitative agreement with the data, albeit with large uncertainties, implying that the predicted total and differential cross sections are of relevance for ongoing and planned experiments.

\end{abstract}


\date{\today}

\maketitle
\section{Introduction}

Baryon-baryon interactions play an important role in \textit{ab initio} studies of nuclear structure, hypernuclei, as well as neutron stars ~\cite{Saha:2004ha,Friedman:2007zza,SchaffnerBielich:2008kb}. In contrast to the nucleon nucleon ($NN$) interaction, where the wealth of experimental data has allowed for the construction of high-precision potentials~\cite{Stoks:1994wp,Wiringa:1994wb,Machleidt:2000ge,Epelbaum:2014sza,Entem:2017gor,Lu:2021gsb}, hyperon-nucleon ($YN$) and hyperon-hyperon ($YY$) interactions still remain poorly constrained because of the lack of high-quality $YN$ and $YY$ scattering  data~\cite{Davis:2005mb,Hashimoto:2006aw,Gal:2016boi}. 

For the $YN$ and $YY$ interactions, the old data were mainly either total cross sections or threshold parameters, which have relatively large uncertainties~\cite{ENGELMANN1966587,Cline:1968zzb,SechiZorn:1969hk,Charlton:1970bv,Kadyk:1971tc,Eisele:1971mk,Hauptman:1977hr}. Only in the past two decades, differential cross sections were measured. With a scintillating fiber block (SCIFI) technique, the KEK-PS E$251$ experiment~\cite{Ahn:1997wa} measured the $\Sigma^{+}p$ differential cross sections in the momentum range of $300\le P_{\Sigma^{+}} \le 600$ MeV$/c$ for two angles $-0.4 \leq \mathrm{cos\theta_{c.m.}} \leq 0.1$ and $0.1 \leq \mathrm{cos\theta_{c.m.}} \leq 0.6$. Afterwards, the KEK-PS E$289$ experiment~\cite{Kondo:2000hn} obtained the differential cross sections of the $\Sigma^-p$ elastic scattering in the momentum range of $400 \le P_{\Sigma^{-}} \le 700$~MeV$/c$. This is the first measurement of the $\Sigma^{-}p$ elastic scattering in the momentum region where the contributions of $P$- and higher partial waves are notable. In 2005, the KEK-PS E289 experiment~\cite{Ahn:2005gb} updated their study of the $\Sigma^{+}p$ scattering in the momentum region of $350 \leq P_{\Sigma^{+}} \leq 750$~MeV$/c$ with three times more data than the KEK-PS E$251$ experiment~\cite{Ahn:1997wa}.  Recently, a new measurement on $\Sigma^{-} p$ scattering with high statistics was performed at the J-PARC Hadron Experimental Facility by the E40 experiment~\cite{Miwa:2021khu}.  Differential cross sections of the $\Sigma^- p$ elastic scattering were extracted with a drastically improved accuracy for the $\Sigma^{-}$ momentum ranging from $470$ to $850$ MeV$/c$. They also performed  the first precise measurement of the differential cross sections of the $\Sigma^-p \to \Lambda n$ reaction in the momentum range of 470-650 MeV$/c$~\cite{J-PARCE40:2021bgw}. The CLAS Collaboration studied the $\Lambda p \rightarrow \Lambda p$ elastic
scattering cross sections in the incident $\Lambda$ momentum range of 0.9–2.0 GeV$/c$, which are the first data on this
reaction since the 1970s~\cite{CLAS:2021gur}. In addition, with high precision correlation techniques, the ALICE Collaboration studied the coupling strength of the $N\Sigma\leftrightarrow N\Lambda$ in the $p\Lambda$ system~\cite{ALICE:2021njx}. The opening of the inelastic $N\Sigma$ channel is clearly visible in the extracted correlation function as a cusp-like structure occurring at a relative momentum of $289$ MeV$/c$.
All these new measurements, though of relatively higher energy, could impose strong constraints on theoretical  hadron-hadron interactions.

In the past, theoretical $YN$ interactions were mainly based on phenomenological models, such as the meson-exchange models by the Nijmegen~\cite{Rijken:1998yy} and J\"ulich ~\cite{Haidenbauer:2005zh} groups. In recent years, lattice QCD ~\cite{Beane:2006gf,Miyamoto:2016hqo,Nemura:2017vjc,Sasaki:2018mzh,Ishii:2018ddn,Doi:2017zov} and chiral effective field theory (ChEFT)~\cite{Savage:1995kv,Korpa:2001au,Hammer:2001ng,Beane:2003yx,Polinder:2006zh,Machleidt:2011zz,Haidenbauer:2013oca,Haidenbauer:2015zqb} have made remarkable progress. Recently, motivated by the successes of the covariant chiral EFT in the study of $NN$ scattering~\cite{Ren:2016jna}, we have extended the covariant ChEFT to the strangeness $S=-1$~\cite{Li:2016paq,Li:2016mln,Song:2018qqm}, $S=-2$~\cite{Li:2018tbt} and  $S=-3,-4$~\cite{Liu:2020uxi} baryon-baryon systems.

It should be noted that most of these $YN$ interactions were only constrained by the low energy $YN$ cross section data. For instance, the low energy constants (LECs) in the $S=-1$ $YN$ interactions were determined by fitting to the 36 low energy data (see Table I) in both the non-relativistic~\cite{Polinder:2006zh,Haidenbauer:2013oca,Haidenbauer:2015zqb} and covariant~\cite{Li:2016paq,Li:2016mln,Song:2018qqm,Liu:2020uxi} ChEFTs. As a result, it remains to be checked whether they are still applicable either for unfitted high energy regions~\cite{CLAS:2021gur}  or for differential cross sections, particularly, those of the J-PARC E40 experiment~\cite{Miwa:2021khu,J-PARCE40:2021bgw}. 
In the present work, based on the leading order covariant ChEFT $YN$ interaction ~\cite{Li:2016paq,Li:2016mln,Song:2018qqm,Li:2018tbt,Liu:2020uxi}, we predict the total and differential cross sections  in the strangeness $S=-1$ sector for the appropriate experimental momentum and compare these results with the latest J-PARC E40 ~\cite{Miwa:2021khu,J-PARCE40:2021bgw}, CLAS~\cite{CLAS:2021gur}, ALICE~\cite{ALICE:2021njx} data, and those from the phenomenological models~\cite{Haidenbauer:2005zh,Rijken:2010zzb,Fujiwara:2001jg} and the heavy baryon ChEFT~\cite{Polinder:2006zh,Haidenbauer:2013oca,Haidenbauer:2019boi}.

It should be stressed that the main purpose of the present study is twofold. First, compared to Ref.~\cite{Li:2016mln}, we present the predictions of the leading order covariant ChEFT for all the $S=-1$ hyperon-nucleon channels up to the laboratory momentum of 900 MeV$/c$, and we  update the estimate of theoretical uncertainties using the method of Refs.~\cite{Epelbaum:2014sza,Epelbaum:2014efa} which has found wide applications in studying the nucleon-nucleon interaction. These results are helpful for planning  future experiments.  Second, by comparing with the latest experimental data from CLAS~\cite{CLAS:2021gur}, and J-PARC~\cite{Miwa:2021khu,J-PARCE40:2021bgw}, particularly the latter, we test the leading-order results, identify discrepancies, and therefore motivate further theoretical studies in  ChEFTs.

This paper is organized as follows. In Sec.~II, we briefly review the covariant ChEFT and explain our strategy to determine the unknown LECs. In Sec.~III we show the numerical results and compare them with available experimental data, followed by a short summary and outlook in Sec.~IV.

\section{Leading order covariant chiral effective field theory }

\begin{figure}[h]
  \centering
  \includegraphics[width=0.2\textwidth]{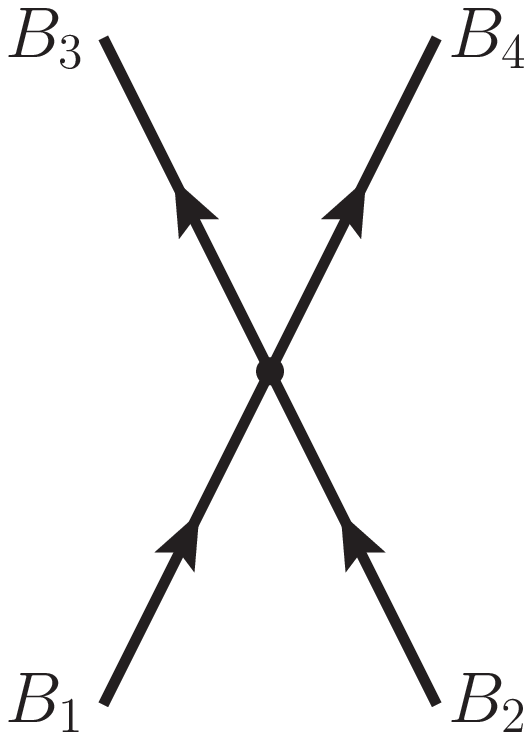} \qquad
  \includegraphics[width=0.2\textwidth]{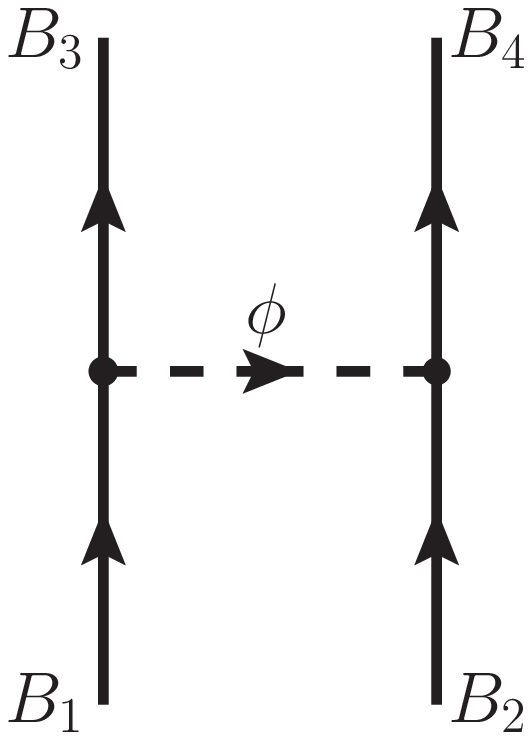}
  \caption{Leading order Feynman diagrams for non derivative four-baryon contact terms and one-meson exchanges.}\label{CTOME}
\end{figure}

In this section, we briefly introduce the covariant ChEFT for the baryon-baryon ($BB$) interaction. At leading order, the $BB$ potentials consist of contributions from non derivative four-baryon contact terms (CT) and one-meson exchanges (OME), as shown in Fig.~\ref{CTOME}. 

The leading-order (LO) Lagrangian for the contact terms is
 \begin{align}\label{CT}
  \mathcal{L}_{\textrm{CT}} = \sum_{i=1}^5\left[\frac{\tilde C_i^1}{2}~\textrm{tr}\left(\bar B_1 \bar B_2 (\Gamma_i B)_2 (\Gamma_i B)_1\right)
  + \frac{\tilde C_i^2}{2}~\textrm{tr}\left(\bar B_1 (\Gamma_i B)_1 \bar B_2 (\Gamma_i B)_2\right)
  + \frac{\tilde C_i^3}{2}~\textrm{tr}\left(\bar B_1 (\Gamma_i B)_1\right)\textrm{tr}\left( \bar B_2 (\Gamma_i B)_2\right)\right],
\end{align}
where $C_i$ ($i=1\ldots 5$) are the LECs that need to be determined by fitting to either experimental or lattice QCD data, and $\Gamma_i$ ($i=1\ldots 5$) are the elements of the Clifford algebra,
\[\Gamma_1=1,\qquad \Gamma_2=\gamma^\mu,\qquad \Gamma_3=\sigma^{\mu\nu},\qquad \Gamma_4=\gamma^\mu\gamma_5,\qquad \Gamma_5=\gamma_5.\]
As discussed in Ref.~\cite{Li:2016mln}, 12 independent combinations of the $15$ LECs survive in the $S = -1$ system, assuming strict SU(3) symmetry. They are collected in Ref.~\cite{Liu:2020uxi}.

To construct the OME potentials, we need the following LO meson-baryon Lagrangian:
\begin{align}\label{LMB1}
  &\mathcal{L}_{MB}^{(1)} =
  \mathrm{tr}\Bigg( \bar B \big(i\gamma_\mu D^\mu - M_B \big)B -\frac{D}{2} \bar B \gamma^\mu\gamma_5\{u_\mu,B\}
  -  \frac{F}{2}\bar{B} \gamma^\mu\gamma_5 [u_\mu,B]\Bigg)\ ,
\end{align}
where $D^\mu B = \partial_\mu B+[\Gamma_\mu,B] $ is the derivative with $\Gamma_\mu$ and $u_\mu$ defined as
\[
  \Gamma_\mu = \frac{1}{2}\left(u^\dag\partial_\mu u + u\partial_\mu u^\dag \right), \quad u_\mu=i(u^\dagger \partial_\mu u-u\partial_\mu u^\dagger)\
\]
with $u^2= U = \exp\left(i\frac{\sqrt{2}\phi}{f_0}\right)$. The values of the coupling constants are $D + F = 1.277$, $F/(F + D) = 0.4$, and the meson decay constant is $f_0=92.2$ MeV$/c$~\cite{ParticleDataGroup:2020ssz}.

From these Lagrangians, one can straightforwardly obtain the contact and  OME potentials. The scattering amplitudes can then be obtained by solving the coupled-channel Kadyshevsky equation~\cite{Kadyshevsky:1967rs},
\begin{align}\label{SEK}
  & T_{\rho\rho'}^{\nu\nu',J}(p',p;\sqrt{s})
  =
   V_{\rho\rho'}^{\nu\nu',J}(p',p)
   +
  \sum_{\rho'',\nu''}\int_0^\infty \frac{dp''p''^2}{(2\pi)^3} \frac{M_{B_{1,\nu''}}M_{B_{2,\nu''}}~ V_{\rho\rho''}^{\nu\nu'',J}(p',p'')~
   T_{\rho''\rho'}^{\nu''\nu',J}(p'',p;\sqrt{s})}{E_{1,\nu''}E_{2,\nu''}
  \left(\sqrt{s}-E_{1,\nu''}-E_{2,\nu''}+i\epsilon\right)},
\end{align}
where $\sqrt{s}$ is the total energy of the two-baryon system in the center-of-mass frame and $E_{n,\nu''}=\sqrt{\mbox{\boldmath $p$}''^{2}+M^{2}_{B_{n,\nu''}}}$, $(n=1,2)$. The labels $\nu,\nu',\nu''$ denote the particle channels, and $\rho,\rho',\rho''$ denote the partial waves. In practice, the potentials in the scattering equation are regularized with an exponential form factor of the following form,
\begin{align}\label{EF}
  f_{\Lambda_F}(p,p') = \exp \left[-\left(\frac{p}{\Lambda_F}\right)^{4}-\left(\frac{p'}{\Lambda_F}\right)^{4}\right].
\end{align}
More details about the covariant ChEFT can be found in Refs.~\cite{Ren:2016jna,Li:2016mln,Li:2016paq,Song:2018qqm,Ren:2018xxd,Li:2018tbt,Bai:2020yml,Liu:2020uxi}.

It has been customary to use the variation of the cutoff as an error estimator (see, e.g., Ref.~\cite{Machleidt:2011zz}. In recent years, it was proposed that one can treat the difference between the optimal results obtained at different orders as the estimate of truncation uncertainties~\cite{Epelbaum:2014sza,Epelbaum:2014efa}. This approach can be briefly described as follows. The expansion parameters for ChEFT read
$$
Q=\operatorname{Max}\left\{\frac{p}{\Lambda_{b}}, \frac{m_{\pi}}{\Lambda_{b}}\right\},
$$
where $p$ is the baryon momentum in the c.m. frame and $\Lambda_{b}$ is the cutoff or the chiral symmetry breaking scale. In our numerical study, $\Lambda_{b}$ is fixed at the optimal cutoff of 600 MeV/$c$. One can estimate the next to leading order (NLO)  truncation uncertainties as
\begin{align}\label{error}
  \Delta^{\mathrm{NLO}}=\operatorname{Max}\left\{Q^{2} \left|\mathcal{O}^{\mathrm{LO}}\right|, Q \left|\mathcal{O}^{\mathrm{LO}}-\mathcal{O}^{\mathrm{NLO}}\right|\right\},  
\end{align}
where $\mathcal{O}$ represent either total cross sections or differential cross sections in our present case. As we did not have the NLO results, we could use either the experimental data as the NLO results (assuming that at NLO, we can fully reproduce the data), or simply use the first term, i.e., $Q^2 |\mathcal{O}^{LO}|$. It should be noted that in our previous works, we have used the cutoff variation to estimate theoretical uncertainties, while in the present work, we use Eq.~(\ref{error}) to estimate truncation uncertainties. One should keep in mind that because we are missing the NLO results, our estimate of theoretical uncertainties can only be trusted for the observables where our LO results can describe reasonably well the data.

\section{Results and discussion}

In this work we study in detail the following $YN$ reactions for which experimental data exist: 1) the coupled-channel reaction $\Lambda p\rightarrow\Lambda p$, $\Sigma^{+}n$, $\Sigma^{0}p$; 2) the coupled-channel reaction $\Sigma^{-}p\rightarrow\Lambda n$, $\Sigma^{0}n$, $\Sigma^{-}p$; and 3) the single-channel reaction $\Sigma^{+}p\rightarrow\Sigma^{+}p$.

\begin{table}[H]
\footnotesize
\centering
 \caption{Experimental $YN$ total cross sections used in the fitting procedure. Momenta are in units of MeV$/c$ and cross sections in mb.}\label{FIT_CS}
\setlength{\tabcolsep}{4pt}
 \begin{tabular}{c|cc|cc|cc|cc|cc|cc}
  \hline
  \hline
 & \multicolumn{2}{c|}{$\Lambda p \rightarrow \Lambda p$~\cite{SechiZorn:1969hk}} & \multicolumn{2}{c|}{$\Lambda p \rightarrow \Lambda p$~\cite{Alexander:1969cx}} &\multicolumn{2}{c|}{$\Sigma^{-} p \rightarrow \Lambda n$~\cite{ENGELMANN1966587}}  & \multicolumn{2}{c|}{$\Sigma^{-} p \rightarrow \Sigma^{0} n$~\cite{ENGELMANN1966587}} & \multicolumn{2}{c|}{$\Sigma^{-}p \rightarrow \Sigma^{-}p$~\cite{Eisele:1971mk}} & \multicolumn{2}{c}{$\Sigma^{+}p \rightarrow \Sigma^{+}p$~\cite{Eisele:1971mk}}   \\
  \hline     
 &  $P_{\rm{lab}}^{\Lambda}$ & $\sigma_{\rm{exp}}$ & $P_{\rm{lab}}^{\Lambda}$ & $\sigma_{\rm{exp}}$ & $P_{\rm{lab}}^{\Sigma^{-}}$ & $\sigma_{\rm{exp}}$ & $P_{\rm{lab}}^{\Sigma^{-}}$ & $\sigma_{\rm{exp}}$ & $P_{\rm{lab}}^{\Sigma^{-}}$ & $\sigma_{\rm{exp}}$ & $P_{\rm{lab}}^{\Sigma^{+}}$ & $\sigma_{\rm{exp}}$ \\
  \hline     
 &  $135 \pm 15$ & $209 \pm 58$ & $145 \pm 25$ & $180 \pm 22$ & $110 \pm 5$ & $174 \pm 47$ & $110 \pm 5$ & $396 \pm 91$ & $\phantom{.5}135 \pm 2.5$ & $184 \pm 52$ &  $145 \pm 5$ & $123 \pm 62$ \\   
 &  $165 \pm 15$ & $177 \pm 38$  & $185 \pm 15$ & $130 \pm 17$ & $120 \pm 5$ & $178 \pm 39$ & $120 \pm 5$ & $159 \pm 43$ & $142.5 \pm 2.5$ & $152 \pm 38$ & $155 \pm 5$ & $104 \pm 30$\\     
 &  $195 \pm 15$ & $153 \pm 27$ & $210 \pm 10$ & $118 \pm 16$ & $130 \pm 5$ & $140 \pm 28$  & $130 \pm 5$ & $157 \pm 34$ & $147.5 \pm 2.5$ & $146 \pm 30$ &  $165 \pm 5$ & $\phantom{1}92 \pm 18$\\     
 &  $225 \pm 15$ & $111 \pm 18$ & $230 \pm 10$ & $101 \pm 12$ & $140 \pm 5$ & $164 \pm 25$ & $140 \pm 5$ & $125 \pm 25$ & $152.5 \pm 2.5$ & $142 \pm 25$ & $175 \pm 5$ & $\phantom{1}81 \pm 12$ \\     
 &  $255 \pm 15$ & $\phantom{1}87 \pm 13$ & $250 \pm 10$ & $\phantom{1}83 \pm 13$ & $150 \pm 5$ & $147 \pm 19$ & $150 \pm 5$ & $111 \pm 19$ & $157.5 \pm 2.5$ & $164 \pm 32$\\     
 &  $300 \pm 30$ & $\phantom{1}46 \pm 11$ & $290 \pm 30$ & $57 \pm 9$ & $160 \pm 5$ & $124 \pm 14$ & $160 \pm 5$ & $115 \pm 16$  & $162.5 \pm 2.5$ & $138 \pm 19$ \\ 
 &  & &  & & & &  & & $167.5 \pm 2.5$ & $113 \pm 16$ \\
\hline
 $\chi^2$ & \multicolumn{2}{c|}{$4.2$} & \multicolumn{2}{c|}{$1.2$} & \multicolumn{2}{c|}{$2.6$} &  \multicolumn{2}{c|}{$6.4$} & \multicolumn{2}{c|}{$2.3$} & \multicolumn{2}{c}{$0.3$}\\
\hline
\multicolumn{12}{c}{$\Sigma^{-}p$ inelastic capture ratio at rest~\cite{deSwart:1963zz}, $r_R = 0.468 \pm 0.010$. \qquad($\chi^2=0.032$)} \\   
 \hline
 \hline
\end{tabular}
 \end{table}
 
In Table~\ref{FIT_CS}, we collect the low-energy data  with $P_\mathrm{lab}\le350$ MeV$/c$, which consist of total cross sections for the following reactions: $\Lambda p \rightarrow \Lambda p$ from Ref.~\cite{SechiZorn:1969hk} (six data points) and Ref.~\cite{Alexander:1969cx} (six data points), $\Sigma^{-}p \rightarrow \Lambda n$~\cite{ENGELMANN1966587} (six data points), $\Sigma^{-}p\rightarrow\Sigma^0n$~\cite{ENGELMANN1966587} (six data points), $\Sigma^{-}p \rightarrow \Sigma^{-}p$~\cite{Eisele:1971mk} (seven data points), $\Sigma^{+}p \rightarrow \Sigma^{+}p$~\cite{Eisele:1971mk} (four data points). In addition to the low energy $YN$ scattering data, the $\Sigma N$ inelastic capture ratio at rest, $r_R$~\cite{deSwart:1963zz}, is also considered. For reference, we also present the corresponding $\chi^2$ obtained with $\Lambda_F=600$MeV$/c$. Note that the $\Sigma^{+}p \rightarrow \Sigma^{+}p$ and $\Sigma^{-}p \rightarrow \Sigma^{-}p$ cross sections  were obtained by incomplete angular coverage $\cos{\theta} \in [-0.5, 0.5]$ experimentally~\cite{Eisele:1971mk}.

\begin{table}[H]
\footnotesize
\centering
 \caption{Experimental $YN$ total cross sections used in the comparison with the theoretical results, but  not used in the fitting procedure. Momenta are in units of MeV$/c$ and cross sections in units of mb.}\label{NON_FIT_CS}
\setlength{\tabcolsep}{4pt}
 \begin{tabular}{cc|cc|cc|cc|cc|cc}
  \hline
  \hline
  \multicolumn{2}{c|}{$\Lambda p \rightarrow \Lambda p$~\cite{Hauptman:1977hr}}  & \multicolumn{2}{c|}{$\Lambda p \rightarrow \Lambda p$~\cite{Kadyk:1971tc}} & \multicolumn{2}{c|}{$\Sigma^{-} p \rightarrow \Lambda n$~\cite{Petschauer:2016tee}} & \multicolumn{2}{c|}{$\Sigma^{-} p \rightarrow \Sigma^{0} n$~\cite{Petschauer:2016tee}} & \multicolumn{2}{c|}{$\Sigma^{-} p \rightarrow \Sigma^{-} p$~\cite{Kondo:2000hn}} & \multicolumn{2}{c}{$\Sigma^{+} p \rightarrow \Sigma^{+} p$~\cite{Ahn:2005gb}}\\
  \hline     
   $P_{\rm{lab}}^{\Lambda}$ & $\sigma_{\rm{exp}}$ & $P_{\rm{lab}}^{\Lambda}$ & $\sigma_{\rm{exp}}$ & $P_{\rm{lab}}^{\Sigma^{-}}$ & $\sigma_{\rm{exp}}$ & $P_{\rm{lab}}^{\Sigma^{-}}$ & $\sigma_{\rm{exp}}$ & $P_{\rm{lab}}^{\Sigma^{-}}$ & $\sigma_{\rm{exp}}$ & $P_{\rm{lab}}^{\Sigma^{+}}$ & $\sigma_{\rm{exp}}$  \\
  \hline
   $350 \pm 50$ & $17.5 \pm 9\phantom{.5}$ & $500 \pm 100$ & $\phantom{1.5}9 \pm 2\phantom{.5}$ & $175 \pm 25$ & $\phantom{.5}59 \pm 13.5$ & $175 \pm 25$ & $\phantom{.}101 \pm 18\phantom{.}$ & $450 \pm 50$ & $19^{+10}_{-5}$ & $400 \pm 50\phantom{0}$ & $74.8^{+29.8}_{-22.5}$\\          
   $450 \pm 50$ & $\phantom{.5}27 \pm 8\phantom{.5}$ & $650 \pm 50\phantom{0}$ & $16.5 \pm 3.5$ & $225 \pm 25$ & $\phantom{.5}60 \pm 12\phantom{.5}$ & $225 \pm 25$ & $\phantom{.5}75 \pm 12\phantom{.}$ & $550 \pm 50$ & $29^{+14}_{-10}$ & $500 \pm 50\phantom{0}$ & $\phantom{.5}15^{+24.3}_{-15.4}$\\         
   $550 \pm 50$ & $\phantom{1.5}7 \pm 4\phantom{.5}$ & $750 \pm 50\phantom{0}$ & $10.5 \pm 2.5$ & $275 \pm 25$ & $\phantom{.5}42 \pm 8\phantom{1.5}$ & $275 \pm 25$ & $\phantom{.5}44 \pm 8\phantom{.5}$ & $650 \pm 50$ & $15^{+23}_{-8}$ & $650 \pm 100$ & $\phantom{.5}15^{+52.2}_{-26.4}$ \\       
   $650 \pm 50$ & $\phantom{1.5}9 \pm 4\phantom{.5}$ & $850 \pm 50\phantom{0}$ & $\phantom{.5}10 \pm 2.5$ &  $325 \pm 25$ & $24.5 \pm 6\phantom{1.5}$ & $325 \pm 25$ & $\phantom{.5}42 \pm 8\phantom{.5}$ &  &  \\      
   $750 \pm 50$ & $\phantom{.5}14 \pm 5\phantom{.5}$ &  &  & $375 \pm 25$ & $\phantom{.5}13 \pm 4\phantom{1.5}$ & $375 \pm 25$ & $\phantom{.5}18 \pm 4.5$ &  &  \\   
   $850 \pm 50$ & $11.5 \pm 3.5$ &  &   & $425 \pm 25$ & $\phantom{.5}42 \pm 7\phantom{1.5}$ & $425 \pm 25$ & $22.5 \pm 4.5$ &  &  \\       
    &  &  &   & $475 \pm 25$ & $27.5 \pm 5.5\phantom{1}$ & $475 \pm 25$ & $18.5 \pm 4.5$&  &  \\     
    &  &  &   & $525 \pm 25$ & $\phantom{.5}14 \pm 4\phantom{1.5}$ & $525 \pm 25$ & $19.5 \pm 4.5$&  &  \\   
    &  &  &   & $575 \pm 25$ & $31.5 \pm 6.5\phantom{1}$ & $575 \pm 25$ & $\phantom{.5}28 \pm 5.5$&  &  \\   
 \hline
 \hline
\end{tabular}
 \end{table} 
 
In Table~\ref{NON_FIT_CS}, we collect the high energy  $\Lambda p\to \Lambda p$, $\Sigma^\pm p\to \Sigma^\pm p$  cross sections for $P_\mathrm{lab}\ge 350$ MeV$/c$ and $\Sigma^- p\to \Lambda n$ and $\Sigma^-p\to \Sigma^0 n$ for $P_\mathrm{lab}\ge 175$ MeV$/c$.  We stress that these data  are not fitted, and therefore they serve as nontrivial tests on the $YN$ interaction by ChEFT.

\subsection{Total cross sections}

\begin{figure}[htpb]
  \centering
  \includegraphics[width=0.8\textwidth]{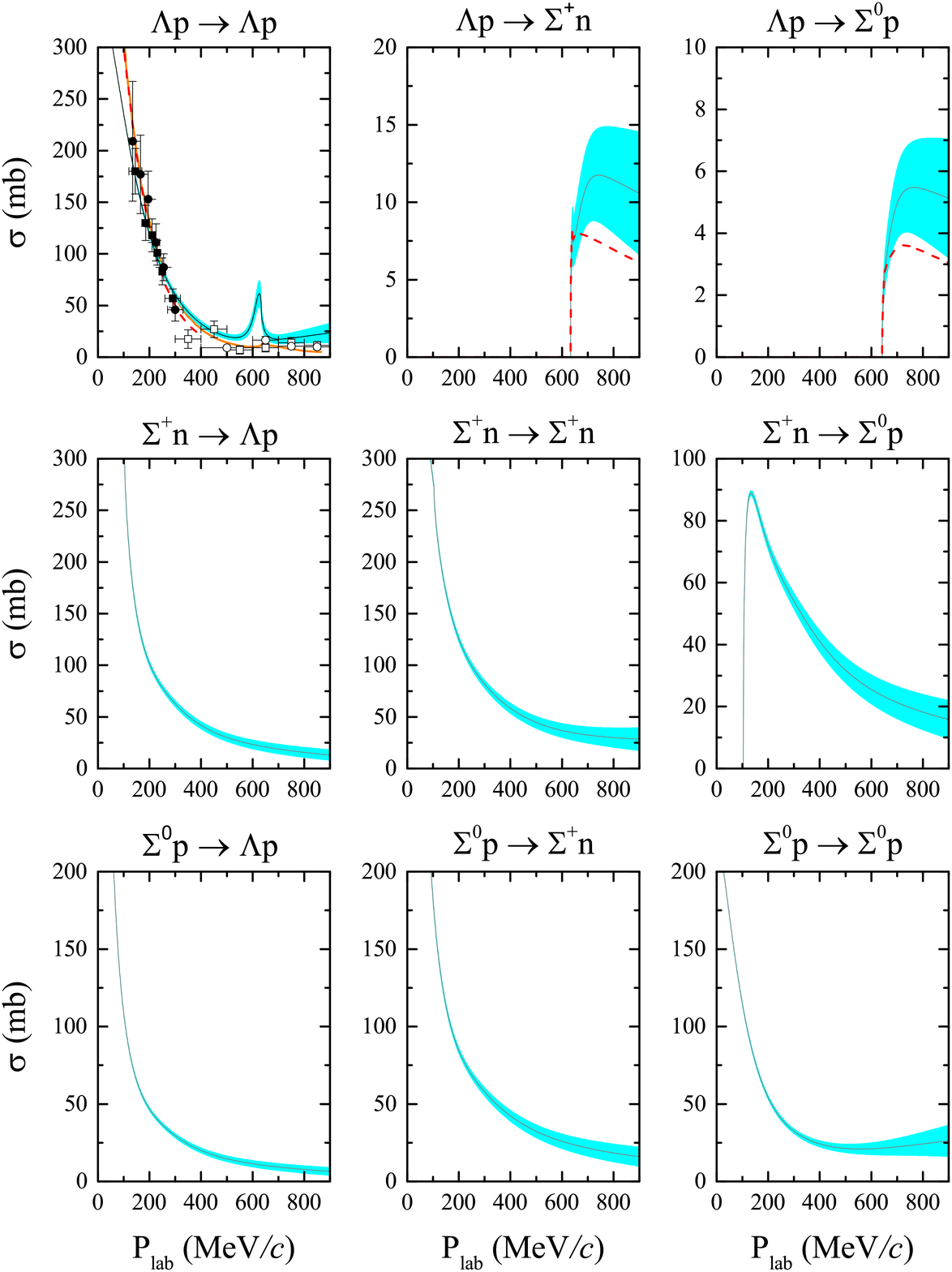}
  \caption{$\Lambda N$ and $\Sigma N$ ($I_3=1/2$) cross sections as functions of the laboratory momentum of the initial hyperon for each reaction as specified at the top of each sub figure. The results are obtained with $\Lambda_F=600$ MeV$/c$, and the bands are theoretical uncertainties estimated using Eq.~(\ref{error}). The experimental data are taken from Sechi-Zorn $et~al$. (filled circles)~\cite{SechiZorn:1969hk}, Alexander $et~al$. (filled squares)~\cite{Alexander:1969cx}, Hauptman $et~al$. (open squares)~\cite{Hauptman:1977hr}, and Kadyk $et~al$. (open circles)~\cite{Kadyk:1971tc}. The additional curves are the theoretical predictions of the meson exchange models, NSC97a (red dashed lines)~\cite{Rijken:1998yy} and J\"ulich'04 (orange solid lines)~\cite{Haidenbauer:2005zh}.}\label{TCS_14}
\end{figure}

\begin{figure}[htpb]
  \centering
  \includegraphics[width=0.8\textwidth]{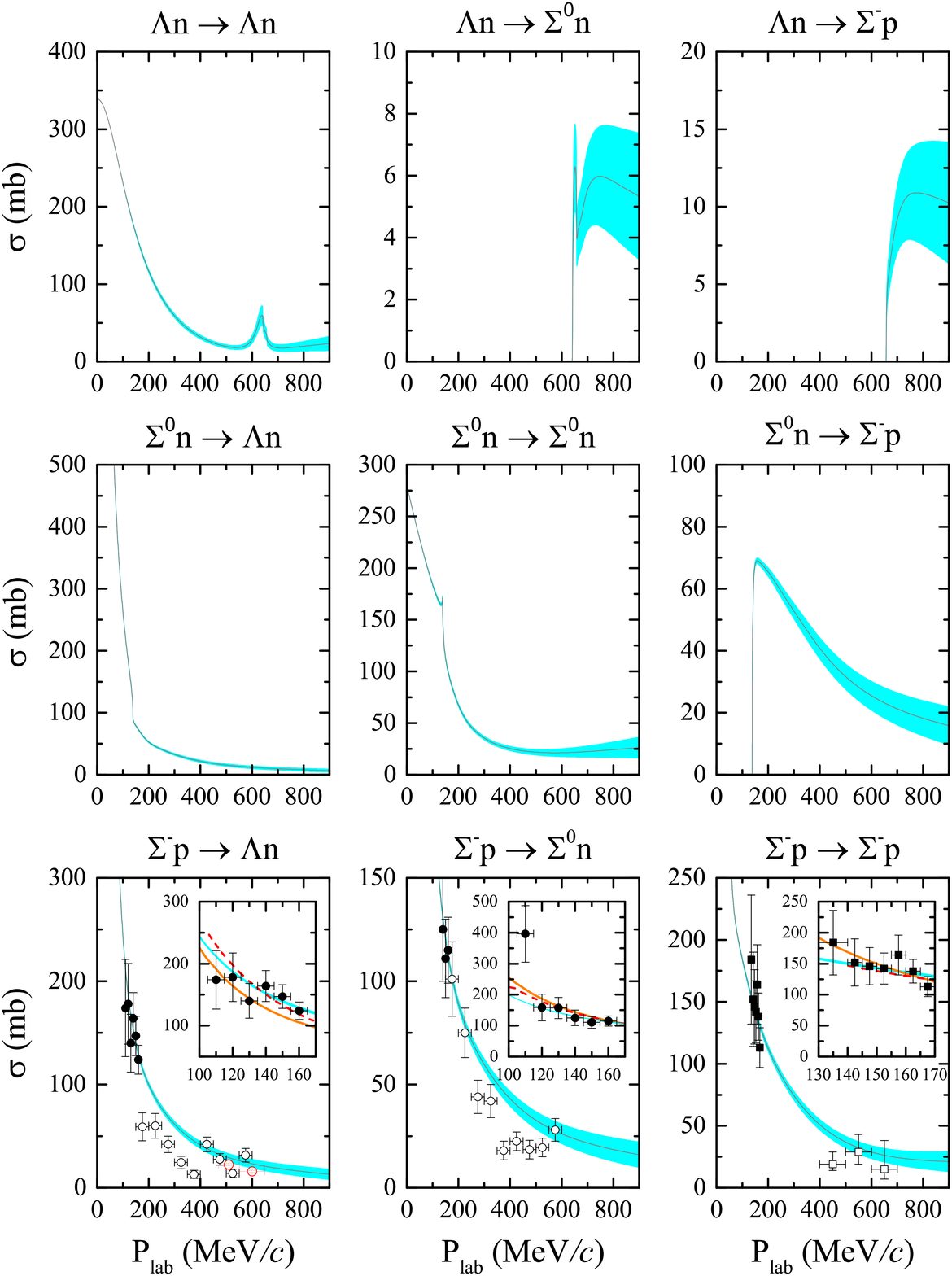}
  \caption{$\Lambda N$ and $\Sigma N$ ($I_3=-1/2$) cross sections as functions of the laboratory momentum of the initial hyperon for each reaction as specified at the top of each subfigure. The results are obtained with $\Lambda_F=600$ MeV$/c$, and the bands are theoretical uncertainties estimated using Eq.~(\ref{error}). The experimental data are taken from Engelmann $et~al$. (filled circles)~\cite{ENGELMANN1966587}, Petschauer:2016tee (open circles)~\cite{Petschauer:2016tee}, Eisele $et~al$. (filled squares).~\cite{Eisele:1971mk}, Kondo $et~al$. (open squares)~\cite{Kondo:2000hn} and Miwa $et~al$. (open red circles)~\cite{J-PARCE40:2021bgw}. The additional curves are the theoretical predictions of the meson exchange models, NSC97a (red dashed lines)~\cite{Rijken:1998yy} and J\"ulich'04 (orange solid lines)~\cite{Haidenbauer:2005zh}.}\label{TCS_15}
\end{figure}

\begin{figure}[htpb]
  \centering
  \includegraphics[width=0.3\textwidth]{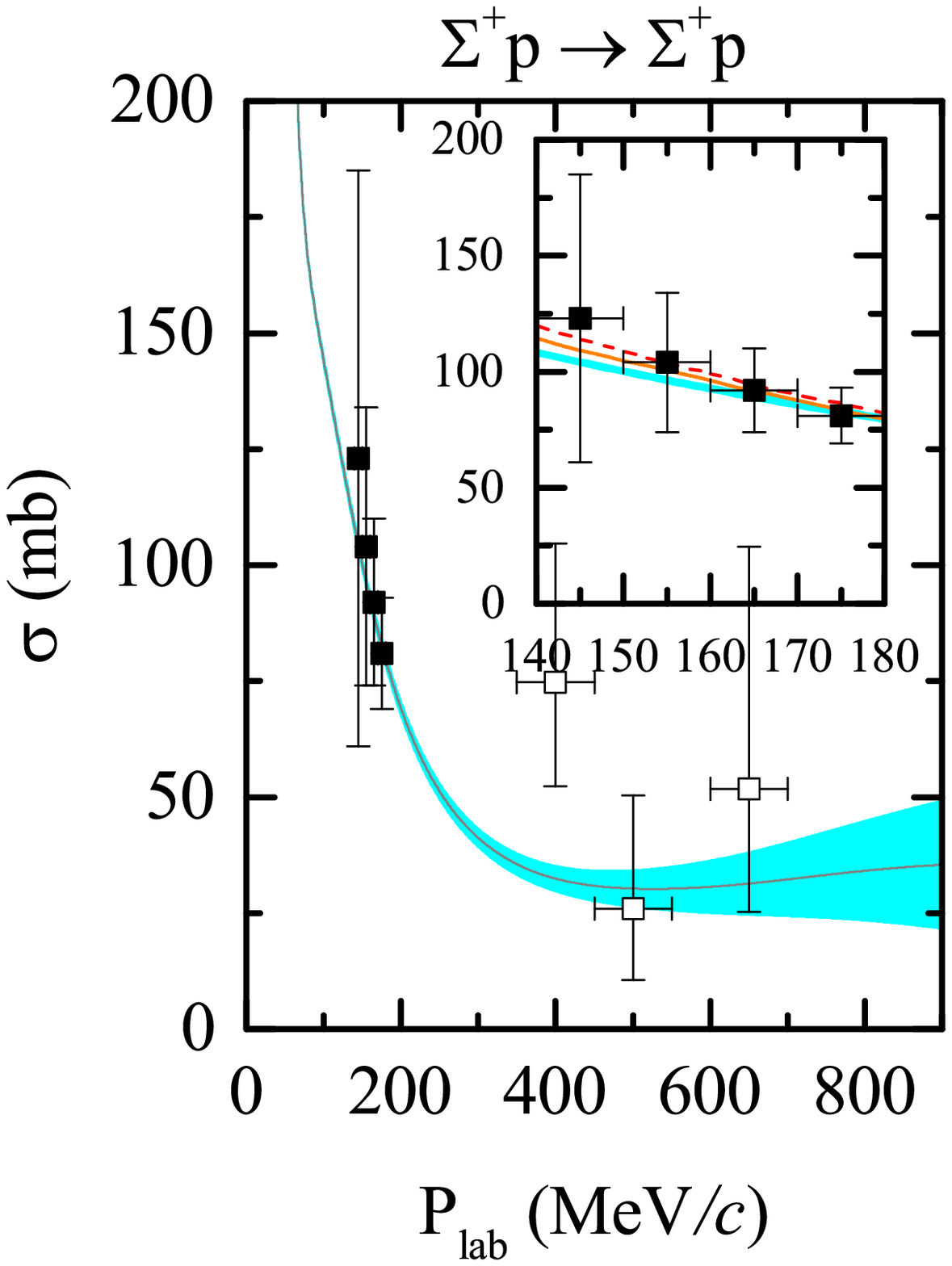}
  \includegraphics[width=0.3\textwidth]{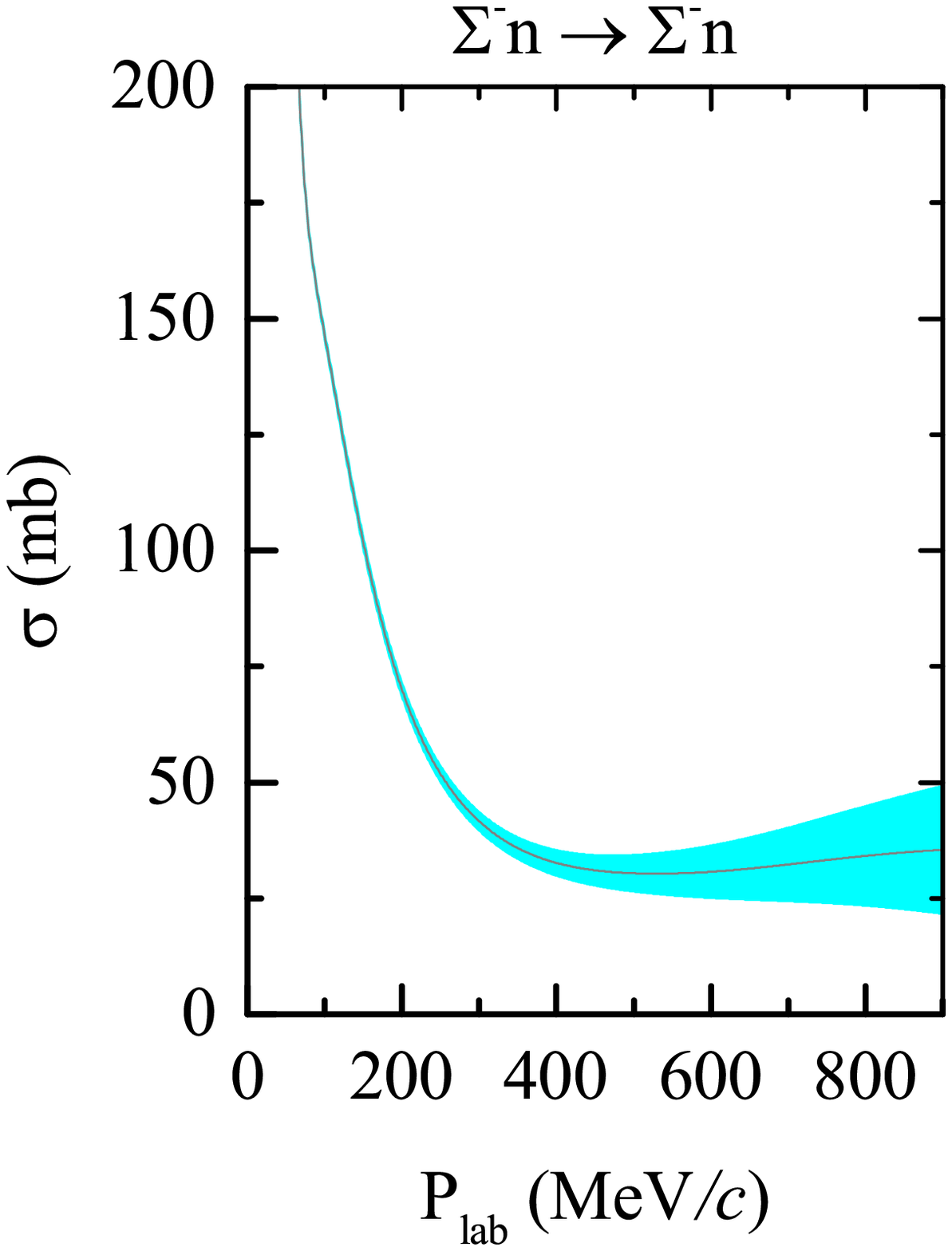}
  \caption{$\Sigma^{+} p$ and $\Sigma^{-} n$ cross sections as functions of the laboratory momentum of the initial hyperon for each reaction as specified at the top of each sub figure. The results are obtained with $\Lambda_F=600$ MeV$/c$, and the bands are theoretical uncertainties estimated using Eq.~(\ref{error}). The experimental data are taken from Eisele $et~al$. (filled squares)~\cite{Eisele:1971mk} and Ahn $et~al$. (open squares)~\cite{Ahn:2005gb}. The additional curves are the theoretical predictions of the meson exchange models, NSC97a (red dashed lines)~\cite{Rijken:1998yy} and J\"ulich'04~\cite{Haidenbauer:2005zh} (orange solid lines).}\label{TCS_13}
\end{figure}

We show the predicted total cross sections for all the $S=-1$ channels below the laboratory momentum about $900$ MeV$/c$ in Figs.~\ref{TCS_14},\ref{TCS_15},\ref{TCS_13}, in comparison with the available experimental data. It should be noted that the covariant ChEFT results were obtained with the optimal cutoff of $600$ MeV$/c$, while the uncertainties are obtained using  Eq.~(\ref{error}). In these figures, the cross section data~\cite{SechiZorn:1969hk,Alexander:1969cx,ENGELMANN1966587,Eisele:1971mk} included in the fitting procedure are denoted by filled symbols, while for the high energy data~\cite{Herndon:1967zz,Kondo:2000hn,Kadyk:1971tc,Hauptman:1977hr,Petschauer:2016tee,Ahn:2005gb} open symbols are used. For the sake of comparison, we also show the results from the J\"ulich'04 model (orange solid lines)~\cite{Haidenbauer:2005zh} and the Nijmegen NSC97f potential (red dashed lines)~\cite{Rijken:1998yy}, if available.  
It is clear that the  ChEFT results agree with the experimental data reasonably well not only at low energies but also at high energies. Note that the high energy results have to be considered as genuine predictions because none of them were included in the fitting procedure and, therefore, the agreement with data demonstrates the predictive power of the covariant ChEFT.  

From Figs.~\ref{TCS_14} and \ref{TCS_15}, we  observe that the $\Lambda p\rightarrow\Lambda p$ and $\Lambda n\rightarrow\Lambda n$  total cross sections show a cusp structure when the $\Sigma^+ n$ and $\Sigma^0 n$ thresholds open. The former shows a pronounced cusp of almost $50$ mb at the $\Sigma^{+} n$ threshold, while the magnitude of the latter is relatively smaller. Because the cusps occur over a very narrow momentum range, it is hard to observe them experimentally. Nonetheless, there is experimental evidence for an enhancement in the $\Lambda p\to \Lambda p$ cross section near the $\Sigma N$ threshold~\cite{Tan:1969jq,Cline:1968zzb,Braun:1977ma,Sims:1971ch,Alexander:1969cx,Eastwood:1971wj,Pigot:1984pe,Siebert:1994jy}, as shown in Fig.~\ref{TCS_14}. 

The cusp structure at the $\Sigma N$ threshold in the total $\Lambda N\to \Lambda N$ cross sections can be traced back to the strong $\Lambda N-\Sigma N~(I=1/2)~^3S_1-{}^3D_1$ coupling induced by the tensor force. This can be checked by examining Fig.~7 of Ref.~\cite{Li:2016mln},   where the $\Lambda N-\Sigma N$  $^3S_1$  phase shifts show a prominent cusp structure at the $\Sigma N$ threshold, while  the cusp in the $^1S_0$ channel is much small.  In fact, a cusp-like structure at the $\Sigma N$ threshold has been observed in the $\Lambda p$ correlation function, which represents the first direct experimental observation of the $\Lambda N$-$\Sigma N$ coupled-channel effect in the $\Lambda p$ system~\cite{ALICE:2021njx}. The study based on the nonrelativistic ChEFT shows that the LO potential predicts a smaller $\Sigma N$ cusp with respect to the NLO potential, while the latter one is more consistent with the experimental data~\cite{ALICE:2021njx}. In other words, the measured correlation function confirms the strength of the coupled-channel $\Lambda N$-$\Sigma N$ interaction in the nonrelativistic NLO potential. By comparing the $\Lambda N$ cross sections at the $\Sigma N$ threshold predicted by the relativistic and nonrelativistic ChEFT~\cite{Haidenbauer:2013oca,Haidenbauer:2021smk}, it is interesting to note that the relativistic LO result is comparable with the nonrelativistic NLO result, which indicates the reliability of the leading order relativistic $\Lambda N$-$\Sigma N$ interaction.

It is necessary to stress that one main purpose of the present work is to  predict cross sections for many coupled channels that have not been measured yet such that they could be checked by future experiments at J-PARC~\cite{Nakada:2019jzs,Miwa:2019ezi}, BEPC~\cite{Yuan:2021yks}, LHC~\cite{Tolos:2020aln}, or HIAF~\cite{Rappold:2020yia}. 

\begin{figure}[htpb]
  \centering
  \includegraphics[width=0.5\textwidth]{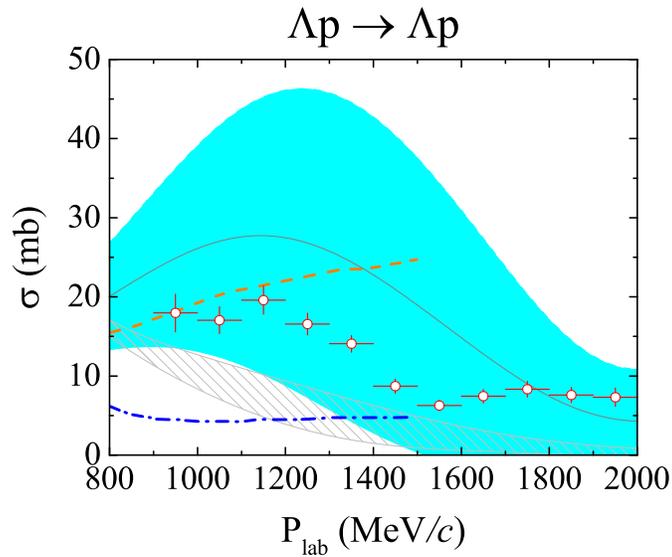}
  \caption{$\Lambda p\to \Lambda p$ cross sections as functions of the laboratory momentum. The results are obtained with $\Lambda_F=600$ MeV$/c$, and the bands are theoretical uncertainties estimated using Eq.~(\ref{error}). For comparison, we also present the results of the leading order HB ChEFT (gray shadow band)~\cite{Polinder:2006zh}. The experimental data are taken from  Rowley $et~al$. (open red circles)~\cite{CLAS:2021gur}. The additional curves are the theoretical predictions of the meson exchange models, NSC97a (orange dashed line)~\cite{Rijken:1998yy} and J\"ulich'04 (blue dashed dotted line)~\cite{Haidenbauer:2005zh}.}\label{TCS_14_high}
\end{figure}

Lately, the CLAS Collaboration reported the first measurement of the  $\Lambda p \rightarrow \Lambda p$ elastic
scattering cross section in the incident $\Lambda$ momentum range of $0.9$–$2.0$ GeV$/c$, which are the first data on this
reaction since the 1970s~\cite{CLAS:2021gur}. Although the momentum range is much larger than what one expects a leading order ChEFT study can cover, it is interesting and instructive to check how they compare with the data. In Fig.~\ref{TCS_14_high}, we compare the covariant ChEFT results with the CLAS data. For the sake of comparison, we also show the results of the J\"ulich model~\cite{Haidenbauer:2005zh}, the Nijmegen model~\cite{Rijken:1998yy}, and those of the LO heavy baryon ChEFT. Clearly, none of them can reproduce the data for such high energies, which are far away from the region where all these models and EFTs were calibrated. Nonetheless, the covariant CHEFT results are not particularly worse either. It is interesting to note that somehow the covariant ChEFT results reach the maximum around the same momentum as the data do. Of course, for such higher energies, one should not trust too much the LO results, and even the NLO results. In addition, as the expansion parameter $Q$ is close to unity when the laboratory momentum reaches 1500 MeV$/c$, the chiral expansion breaks down for this and higher energies. On the other hard, it is gratifying to see that the predictions at least provide a reasonable estimate of order of magnitude of the data. We leave a more careful and systematic study to a future work.

\subsection{$\Sigma^{+}p$ and $\Sigma^{-}p$ differential cross sections }

\begin{figure}[htpb]
  \centering
  \includegraphics[width=0.8\textwidth]{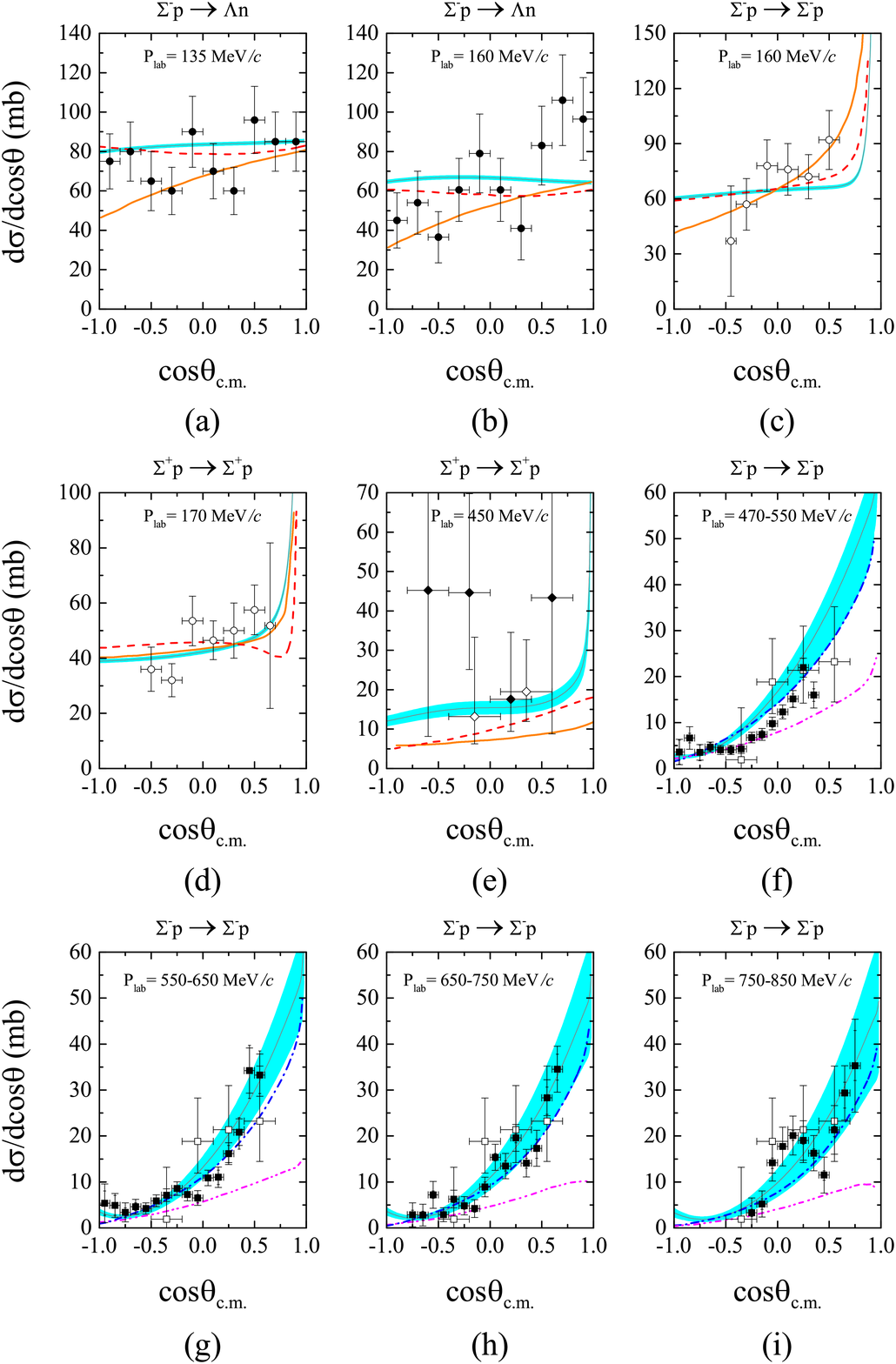}
  \caption{$\Lambda N$ and $\Sigma N$ differential cross section $d\sigma/d \cos{\theta}$ as a function of $\cos{\theta}$, where $\theta$ is the c.m. scattering angle, for various  $P_{\textrm{lab}}$. The results are obtained with $\Lambda_F=600$ MeV$/c$, and the bands are theoretical uncertainties estimated using Eq.~(\ref{error}). In (a), (b), (c), (d), and (e), the experimental data are taken from Engelmann $et~al$. (filled circles)~\cite{ENGELMANN1966587}, Eisele $et~al$. (open circles)~\cite{Eisele:1971mk}, and Ahn $et~al$. (filled diamonds and open diamonds)~\cite{Ahn:1997wa,Ahn:2005gb}. In (f), (g), (h), and (i), the experimental data  (filled squares) are taken from the J-PARC E40 experiment~\cite{Miwa:2021khu} for the $\Sigma^{-}p$ elastic channel of the laboratory momentum from $470$ to $850$ MeV$/c$. The open squares are the experimental data obtained in the KEK-PS E289 experiment~\cite{Kondo:2000hn}. The additional curves are the theoretical predictions of the meson exchange models, NSC97a (red dashed lines)~\cite{Rijken:1998yy} and J\"ulich'04 (orange solid lines)~\cite{Haidenbauer:2005zh}, and ESC08 (magenta dash-dot-dotted lines)~\cite{Rijken:2010zzb}, and the quark cluster model fss2 (navy dash-dotted lines)~\cite{Fujiwara:2006yh}. 
  }\label{DCS_bands}
\end{figure}

In Fig.~\ref{DCS_bands}, we show the predicted differential cross sections below $P_{\rm{lab}}\leq 900$ MeV$/c$ in comparison with the available data~\cite{ENGELMANN1966587,Eisele:1971mk,Ahn:1997wa,Kondo:2000hn,Ahn:2005gb,Miwa:2021khu}, which were not included in the fitting procedure. Here, the experimental measurements are: (a) $\Sigma^{-}p \rightarrow \Lambda n$ differential cross sections at $P_{\Sigma^{-}} = 135$ MeV$/c$ and $P_{\Sigma^{-}} = 160$ MeV$/c$~\cite{ENGELMANN1966587}, (b) $\Sigma^{+}p$ elastic scattering at $P_{\Sigma^{+}} = 170$ MeV$/c$~\cite{Eisele:1971mk} and $P_{\Sigma^{+}} = 450$ MeV$/c$~\cite{Ahn:1997wa,Ahn:2005gb}, (c) $\Sigma^{-}p$ elastic scattering at $P_{\Sigma^{-}} = 160$ MeV$/c$~\cite{Eisele:1971mk}, $P_{\Sigma^{-}} = 400$-$700$ MeV$/c$~\cite{Kondo:2000hn}, and $\Sigma^{-}p$ elastic scattering at $P_{\Sigma^{-}} = 470$-$850$ MeV$/c$~\cite{Miwa:2021khu}. For comparison, we also plot the $\Sigma^{-}p$ elastic scattering differential cross sections from the one boson exchange model (J\"ulich 04)~\cite{Haidenbauer:2005zh},  the quark-cluster model of the Kyoto–Niigata group (fss2)~\cite{Fujiwara:2006yh}, and the meson exchange model from the Nijmegen group (ESC08c)~\cite{Rijken:2010zzb}. 

 As can be seen from Fig.~\ref{DCS_bands}, the flat theoretical differential cross sections for $\Sigma^{-}p \rightarrow \Lambda n$ at $P_{\textrm{lab}}=135$ MeV$/c$ and $P_{\textrm{lab}}=160$ MeV$/c$ are in reasonable agreement with the  data~\cite{ENGELMANN1966587} within uncertainties. One should note that the experimental data shown in Fig.~\ref{DCS_bands}(a$-$e) are  averages over different momentum intervals. Specifically, for $\Sigma^{-}p \rightarrow \Lambda n$ the data are averages over the intervals $100\leq P_{\Sigma^{-}}\leq 170$ MeV$/c$ and $150\leq P_{\Sigma^{-}}\leq 170$ MeV$/c$~\cite{ENGELMANN1966587}, respectively.  In view of the large experimental uncertainties we refrain here from averaging our theoretical results and, following common practice, present our predictions at the central value of the momenta. The same is also true for the data of Ref.~\cite{Eisele:1971mk} which represent averages over $150\leq P_{\Sigma^{-}}\leq 170$ MeV$/c$ for $\Sigma^{-}p \rightarrow \Sigma^{-}p$ and $160\leq P_{\Sigma^{+}}\leq 180$ MeV$/c$ for $\Sigma^{+}p \rightarrow \Sigma^{+}p$, respectively. 

In the higher energy region, the  experimental differential cross sections  for $\Sigma^{-}p \rightarrow \Sigma^{-}p$ are averages  over $400\leq P_{\Sigma^{-}}\leq 700$~MeV$/c$~\cite{Kondo:2000hn}, while those for $\Sigma^{+}p \rightarrow \Sigma^{+}p$ are averages over $300\leq P_{\Sigma^{+}}\leq 600$~MeV$/c$~\cite{Ahn:1997wa} and $350\leq P_{\Sigma^{+}}\leq 750$~MeV$/c$~\cite{Ahn:2005gb}. The predictions at the central momenta are depicted in  Fig.~\ref{DCS_bands}(f$-$i). The J-PARC E40 experiment measured the differential cross sections of $\Sigma^{-}p\to\Sigma^-p$ at four-momentum intervals, i.e., $470$-$550$~MeV$/c$, $550$-$650$~MeV$/c$, $650$-$750$~MeV$/c$, and $750$-$850$~MeV$/c$ in~\cite{Miwa:2021khu}. Clearly, the covariant ChEFT results agree with the E40 results reasonably well. The quark cluster model can also describe the experimental data quite well but not those of ESC08c.  As a result, one can conclude that the latest  $\Sigma^{-}p\to\Sigma^- p$ differential cross section data indeed impose strong constraints on the theoretical $BB$ interactions.~\footnote{It should be mentioned that in Ref.~\cite{Song:2018qqm}, it was shown that the NSC97f  $\Sigma N$ phase shifts for $^3S_1$ with $I=3/2$ are  not consistent with the lattice QCD data.}

\begin{figure}[htpb]
  \centering
  \includegraphics[width=0.9\textwidth]{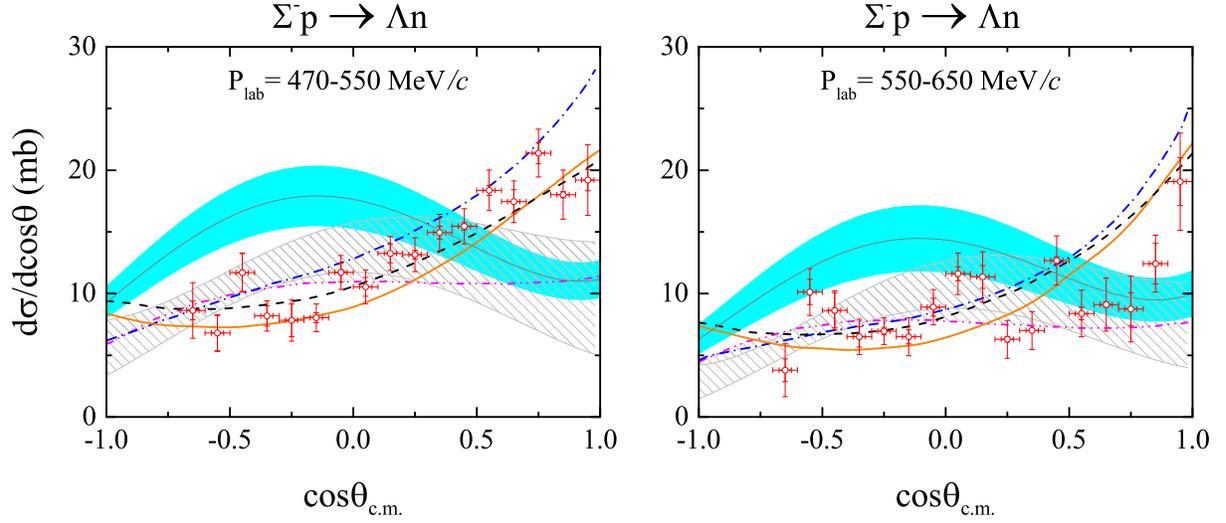}
  \caption{$\Sigma^{-}p \longrightarrow \Lambda n$ differential cross section $d\sigma/d \cos{\theta}$ as a function of $\cos{\theta}$, where $\theta$ is the c.m. scattering angle, for various  $P_{\textrm{lab}}$. The results are obtained with $\Lambda_F=600$ MeV$/c$, and the bands are theoretical uncertainties estimated using Eq.~(\ref{error}). For comparison, we also present the results of the leading order HB ChEFT (gray shadow bands). The  experimental data (open red circles) are taken from the J-PARC E40 experiment~\cite{J-PARCE40:2021bgw} for the $\Sigma^{-}p\longrightarrow \Lambda n$ reaction of the laboratory momentum from $470$ to $650$ MeV$/c$. The additional curves are the theoretical predictions of the meson exchange models, ESC08 (magenta dash-dot-dotted lines)~\cite{Rijken:2010zzb}, and the quark cluster model fss2 (navy dash-dotted lines)~\cite{Fujiwara:2006yh}. The solid orange and black dashed lines represent the results of the HB ChEFT  model, NLO13~\cite{Haidenbauer:2013oca} and NLO19~\cite{Haidenbauer:2019boi}, respectively.
  }\label{DCS_bands_New}
\end{figure}

In Ref.~\cite{J-PARCE40:2021bgw}, the J-PARC E40 Collaboration reported the inelastic differential cross sections of $\Sigma^{-}p\to\Lambda n$ for two momentum intervals, i.e., $470$-$550$~MeV$/c$ and $550$-$650$~MeV$/c$. They are compared with predictions of the LO covariant ChEFT, those of the LO~\cite{Polinder:2006zh} and NLO~\cite{Haidenbauer:2013oca,Haidenbauer:2019boi} HB ChEFT, as well as those of the quark cluster model~\cite{Rijken:2010zzb,Fujiwara:2006yh} in  Fig.~\ref{DCS_bands_New}. It is clear that at LO neither the covariant  ChEFT nor the HB ChEFT  can reproduce the data at a quantitative level, but the fss2  and NLO HB ChEFT results are in much better agreement with the data. Considering that the LO covariant ChEFT can describe the elastic channel reasonably well (see Fig.~\ref{DCS_bands}), the poor performance for the inelastic channel is a bit unexpected. Nonetheless, the good performance of the NLO HB ChEFT clearly suggests that one need to perform higher order studies for this inelastic channel.

\section{Conclusion and outlook}

We predicted the total  $S=-1$ $YN$ cross sections in a wide energy range up to $P_\mathrm{lab}=900$ MeV$/c$ and for all the allowed channels based on the covariant chiral effective field theory. In particular, we showed that the predicted  $\Sigma^{+}p$ and $\Sigma^{-}p$ differential cross sections are in reasonable agreement with the latest J-PARC E40 data. The comparison with other models showed that the differential cross sections can help better constrain  theoretical models.  It should be noted that although  the qualitative agreement with data  supports the purpose of the present work, i.e., providing predictions that can be used as (rough) guidance to plan future experiments, the comparison with either the J-PARC $\Sigma^- p\to\Lambda n$ differential cross sections or the CLAS $\Lambda p$ cross section data show that higher order ChEFT studies are needed. On the other hand, the present study showed clearly how the measurement of different cross sections can help better constrain theoretical descriptions of baryon-baryon interactions, which otherwise cannot be distinguished between each other using only the total cross section data.

The hyperon-nucleon potentials are receiving much attention in recent years because they play an important role in our understanding of  hypernuclear physics and dense neutron stars. In addition, there are ongoing and planned experimental efforts to measure them at facilities such as J-PARC, BEPC, LHC, JLab, and HIAF.  In recent years, lattice QCD simulations have also greatly advanced our understanding of the hyperon-nucleon interactions and will achieve more in the near future. We hope that the results presented in this  work will stimulate more future experimental, theoretical, and lattice QCD studies.

\section{Acknowledgements}
This work was partly supported by the National Natural Science Foundation of China (NSFC) under Grants No. 11975041, 11735003, and 11961141004.

\bibliography{refs.bib}
\end{document}